# Training Hybrid Neural Networks with Multimode Optical Nonlinearities Using Digital Twins

Ilker Oguz*, Louis J. E. Suter*, Jih-Liang Hsieh, Mustafa Yildirim, Niyazi Ulas Dinc, Christophe Moser, Demetri Psaltis

*EPFL, Institute of Electrical and Micro Engineering, 1015 Lausanne, Switzerland*

*Equal contribution

## Abstract

The ability to train ever-larger neural networks brings artificial intelligence to the forefront of scientific and technical discoveries. However, their exponentially increasing size creates a proportionally greater demand for energy and computational hardware. Incorporating complex physical events in networks as fixed, efficient computation modules can address this demand by decreasing the complexity of trainable layers. Here, we utilize ultrashort pulse propagation in multimode fibers, which perform large-scale nonlinear transformations, for this purpose. Training the hybrid architecture is achieved through a neural model that differentiably approximates the optical system. The training algorithm updates the neural simulator and backpropagates the error signal over this proxy to optimize layers preceding the optical one. Our experimental results achieve state-of-the-art image classification accuracies and simulation fidelity. Moreover, the framework demonstrates exceptional resilience to experimental drifts. By integrating low-energy physical systems into neural networks, this approach enables scalable, energy-efficient AI models with significantly reduced computational demands.

## Introduction

Modern artificial intelligence (AI) models are currently driving a transformation across a wide variety of fields including science, industry, healthcare, and education with their unprecedented ability to learn an immense scale of information and solve complex problems. These capabilities have been mainly enabled by two key factors: algorithmic breakthroughs in training neural networks with increasingly larger numbers of parameters and datasets, and the advent of massively parallelized digital hardware to execute these algorithms. So far, employing larger neural networks constantly provided higher AI accuracies in various tasks such as language or image processing, given correspondingly large training sets [1]. The observation of this trend has also led to state-of-the-models with exponentially larger number of parameters, nearly doubling every year [2]. However, this rapid increase in model sizes, coupled with the broader adoption of AI, has created a compounding effect, doubling this technology's overall energy consumption every 100 days [3] and resulting in a correspondingly larger environmental impact due to carbon emissions [4].

A promising approach to performing complex tasks without increasing the number of trainable parameters is the addition of random features, where a subset of weights is randomly assigned instead of trained [5], [6]. The effectiveness of this methods have been demonstrated within architectures such as deep convolutional networks [7], transformers [8], [9], and graph neural networks [10]. Despite their potential, random features are not universally integrated into modern neural networks, because they still require significant computation contributing to energy consumption and latency on conventional hardware. To address this challenge, low-power, high-speed, and high-dimensional physical systems offer a promising alternative for efficiently generating large numbers of fixed weights [11], [12], [13], [14]. While fixed physical layers can be used as initial preprocessing in a straightforward manner by training the following digital layers conventionally, adding programmable layers before the physical layer is more challenging, especially for highly complex, nonlinear systems. One such complex system is the propagation of pulses in MMFs, where tight confinement of light over long distances can create nonlinear modal couplings with low power in addition to the linear transformation due to propagation [15], [16]. Backpropagation-free methods, such as genetic algorithms [17], surrogate optimization [18] and local learning [19], have shown promise in significantly reducing parameter counts while achieving competitive accuracy by treating optical systems as fixed black boxes. Nevertheless, further research is needed to scale these approaches for more demanding applications requiring large number of parameters.

On the other hand, to be able to use the error backpropagation algorithm, which successfully trains today's AI models reaching trillions of weights, accurate knowledge of each layer's analytical representation is required, yet this is challenging to achieve with complex physical layers, as shown in Fig. 1a. Instead, approximating their functionality with neural networks trained with examples from the physical system allowed different applications with intricate optical systems including telecommunication [20], femtosecond laser inscription[21], image acquisition [22], and projection [23] alongside training physical AI models [13].

Even if a model can predict the behavior of a subset of all possible samples, it is highly challenging to fully generalize efficient approximators to high-dimensional nonlinear systems such as pulse propagation in multimode fibers. Moreover as in many analog information processing systems [24], [25]drift in the compute characteristics is eventually unavoidable in MMF-based systems as well.

Here, we demonstrate an approach that can introduce nonlinear optical layers to neural networks in a scalable manner and can train them effectively while being resilient to the deteriorating effects mentioned. To achieve this goal, we synergistically update a digital twin of the optical system throughout the training process and approximate the Jacobian matrix of the system, which holds the partial derivatives between its output and input channels. This way, the layers preceding the physical ones can receive feedback on the effect of their weights on the output and can be updated, as indicated in Fig. 1b. Approximation of the gradients over the physical layers allows for the utilization of them in multiple locations, creating large-scale networks with only a small portion of the weights trainable and digital as in Fig. 1c, while the forward process can fully benefit from the optical weights without any additional digital operations during the inference phase. Our experimental results, obtained with a combination of digital and physical layers in the 3-layer system shown in Fig. 2a, quantitatively demonstrate the effectiveness of the method for AI tasks in terms of accuracy, resiliency to drifts, and simulation fidelity.

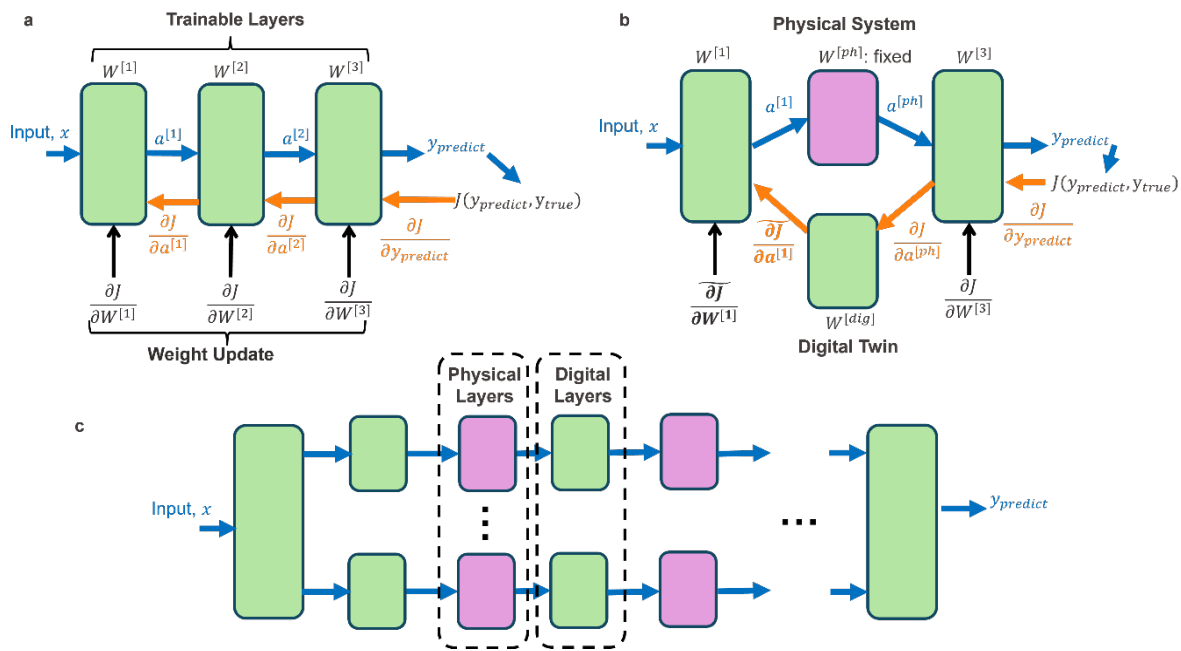

**Figure 1 The information flow during conventional training of neural networks, and training scheme with physical layers. a** In fully digital neural networks, an analytical model of the functionality of each layer is precisely known and their derivatives are used in the calculation of the updates to the weights to minimize error. **b** Physical layers can be included in the forward computation in neural networks while their digital twin serves as a differentiable function to replace them in the backpropagation step. The data acquired in the forward pass serves as training dataset that constantly refines the digital twin. **c** Gradient approximation with the digital twin allows for large scale hybrid networks containing physical layers in different locations

# Methods and Results

## Guiding Model Training with the Optical Layer Twin

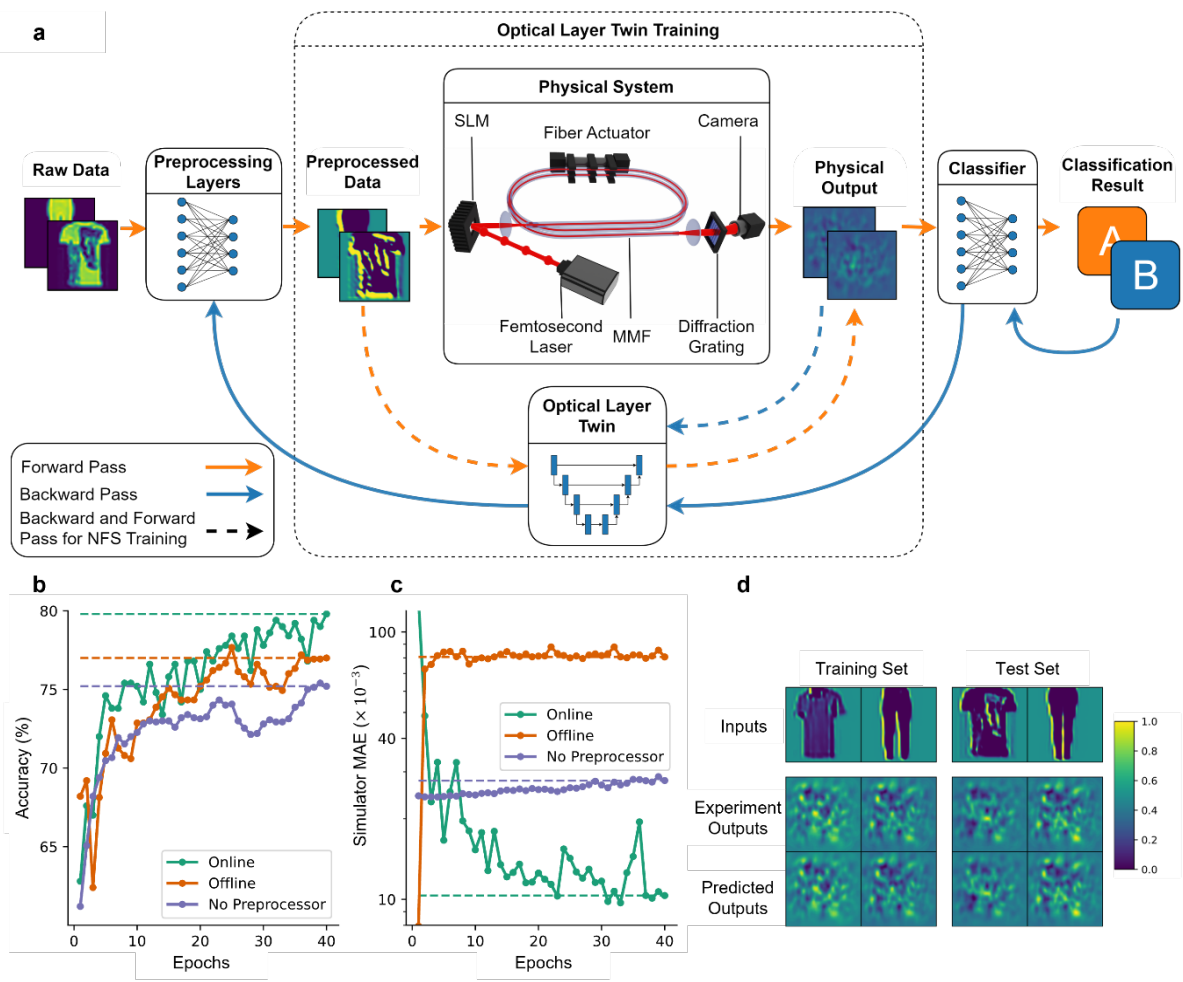

**Figure 2 The architecture, algorithm and outcomes of training of a neural network with a physical layer. a** The data flow in the network through digital and physical layers both in forward and error backpropagation steps. The physical layer consists of an optical system based on the propagation of spatially modulated ultrashort pulses in a multimode fiber. **b** The progress of test accuracy over training epochs for different learning structures on Fashion-MNIST dataset. **c** The mean absolute error (MAE) between Optical Layer Twin (OLT) predictions and experimentally obtained intensity patterns for image in the test set for different training procedures. **d** Some intensity distribution samples from OLT and experimental system for the same input images.

Our approach leverages a digital twin, named Optical Layer Twin (OLT), to approximate the physical response of propagation through the MMF, a high-dimensional nonlinear system. The propagation of spatially phase-modulated ultrashort pulses (10 ps long at 1030 nm with 125 kHz repetition rate, 12.6 mW fiber-coupled average power, generated by Amplitude Laser, Satsuma) in a multimode fiber (OFS, bend-insensitive OM2, 50 μm core diameter, 0.20 NA, 5m long) constitutes the physical event in the optical system. As shown in Fig. 2a, the input information enters the system as a 2-dimensional phase distribution on the spatial light modulator (SLM) (Meadowlark HSP1920), which determines the modes coupled into the fiber. The propagation of the modulated beam through the silica core of the fiber induces spatiotemporal nonlinearities due to $\chi^{(3)}$ light-matter interactions such as cross-phase modulation and four wave mixing, remapping the initial field distribution to a completely different pattern, which is measured finally on a camera. Optical power coupled to the MMF is selected to be the optimal level for obtaining rich nonlinearities without impairing the performance with beam clean-up effects as previously established in the literature [18]. Appendix Note 1 explains further these nonlinear propagation dynamics. The fiber actuator can change the physical conformation of the fiber controllably and can serve as a source of reproducible perturbation of the physical system. Before reaching the camera, a diffraction grating disperses the beam spatially separating its wavelength components. This enriches the transformation between the input phase and the output intensity patterns by distributing the spectral effects to camera pixels.

The physical system provides a large set of nonlinear connections to the overall neural network model, which can learn to tackle various machine learning problems through trainable initial and final layers. In the demonstration shown in Fig. 2, the initial preprocessing layer is a single convolutional layer, while the final classifier is a fully connected layer. During training, OLT accompanies the forward computation path of the model, including the physical system, by approximating the physical response with a multilayer neural network which is a differentiable function for the error signal's backward propagation. OLT itself is a standalone neural network that predicts the output of the physical system given the phase modulation pattern on the SLM. Since incorporating the OLT into the training loop enables updating the weights of the layers preceding the physical model with gradients from the downstream task, it does not add any extra overhead to the forward information flow in the model. Consequently, the computational complexity of inference operations after training remains unchanged. Furthermore, the OLT is continuously updated with experimental data generated during operation to achieve the highest approximation fidelity. We refer to this mode of operation as "Online Learning" in this paper and compare it with an approach where the OLT is pre-configured and kept fixed throughout training, termed "Offline Learning." In Fig 2a, the operations shown with solid lines take place in both approaches, whereas online learning also applies update steps indicated with dashed lines. The details of these training algorithms are provided in Appendix Note 2.

In scenarios where the preprocessing layers are more complex and its weights significantly alter the input data sent to the physical system, online learning becomes increasingly crucial. This is because a fixed OLT, trained on a limited or specific dataset, may fail to generalize to new input distributions introduced by updated preprocessing weights. The wide gap between the dataset used to train the OLT and the new, altered input samples it processes later can reduce the fidelity of the OLT, leading to imprecise updates to the upstream layers in training.

We compared computational performances of the two learning algorithms on a subset of Fashion-MNIST dataset [26] with only 1500 training samples, also including a simpler approach that processes raw data directly with the physical system and classifies its outputs without any preprocessing. As shown in Fig. 2b, adding a preprocessing layer and training with offline learning improves task test accuracy from 75% to 77%. When the training algorithm keeps updating the OLT throughout training, a better performing preprocessor brings the accuracy to 80%. The fidelity of OLT to the actual experiment, visualized in Fig. 2c, explains these differences. Without a preprocessing layer, the slight increase in error stems directly from slow drifts in the experimental system over time, since the same raw data and a pre-trained, fixed OLT is utilized In the offline learning experiment, the error level rises rapidly during the initial epochs, as changes in the preprocessing layer alter the distribution of inputs to the physical system, which the fixed OLT cannot accommodate and MAE reaches nearly $10^{-1}$. Here epoch means one pass over the entire dataset during training. In contrast, with online learning, the OLT tracks these distributional changes, ultimately achieving high precision in predicting experimental outputs. The precision level reaches a normalized MAE of $1.03 \times 10^{-2}$, meaning that signal-to-noise (SNR) ratio of the OLT predictions is $96.8$. The examples in Fig. 2d, taken from the final epoch of the online training procedure, demonstrate the effectiveness of the OLT, with nearly visually identical speckle patterns between the experimental outputs and the blind predictions of the neural simulator. Remarkably, the OLT achieves these accurate predictions with a latency of 30 ms, while an analytical simulation of the wave propagation dynamics, using the multimode nonlinear Schrödinger equation for the same fiber, would take approximately 500 seconds on a GPU, even when limited to only 15 of the fiber's 240 modes [27].

# Learning and Computing with a Physical System under Dynamic Perturbation

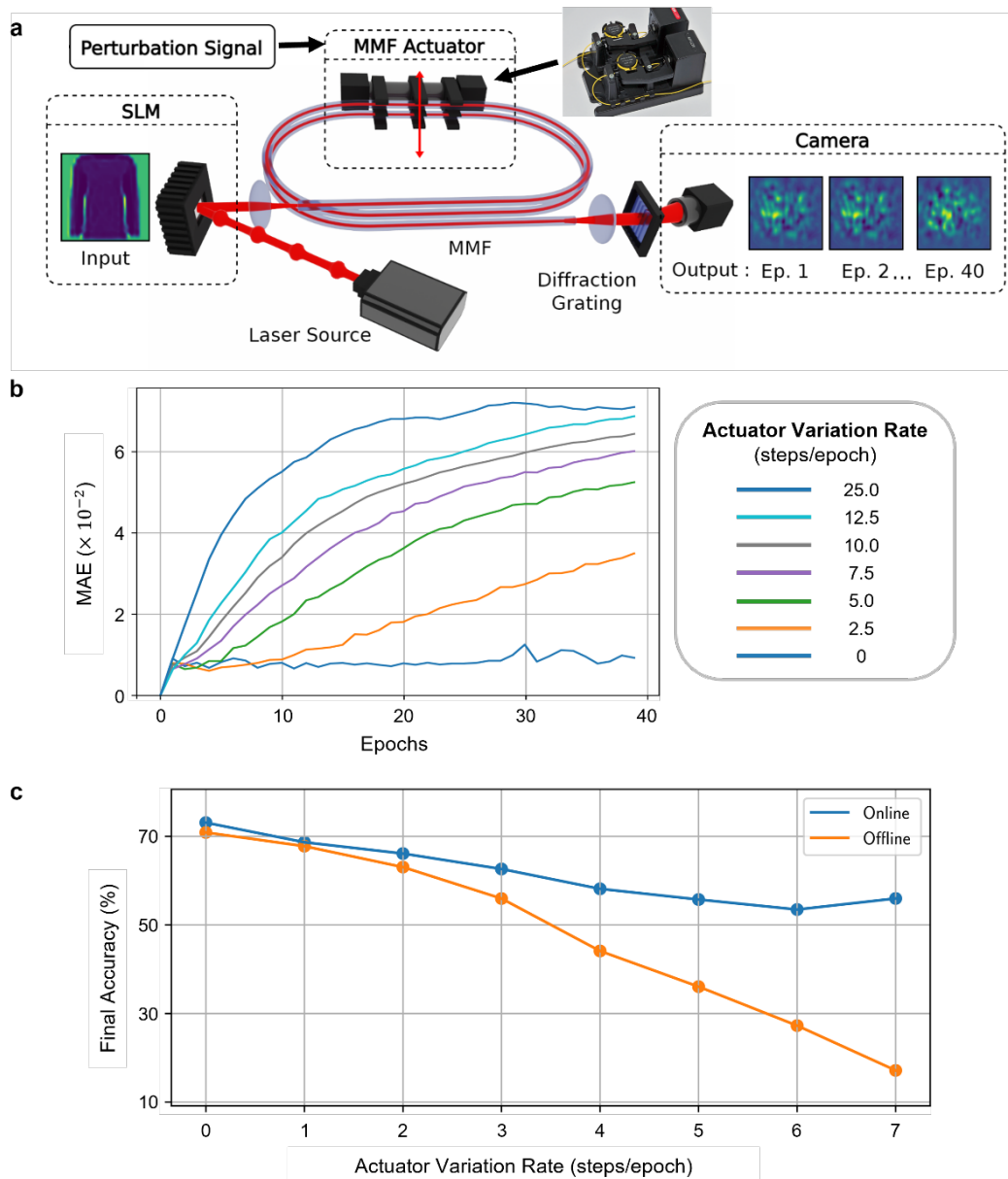

**Figure 3 The method for applying dynamic perturbations to a multimode fiber based optical system, their effects on the output statistics and accuracy of the overall network. a** A mechanical actuator is added to bend the MMF in the setup shown in Fig. 2a, introducing dynamic perturbations. The effects are observed through changes in output speckles at each epoch for the same input pattern. **b** Comparison of the dataset's outputs in the first epoch versus subsequent epochs, quantified using the average MAE at different actuator variation rates. **c** Dependency of final test accuracies on the perturbation rate, shown for offline and online learning schemes.

As in various analog information processing systems, gradual small changes in the working characteristics of optical physical layers can accumulate and significantly impact final predictions. To investigate these effects and their impact on the neural network performance in an accelerated manner, a mechanical actuator is incorporated into the optical system shown in Fig. 2a. The mechanical actuator (Thorlabs MPC320), originally designed to create stress-induced birefringence in single-mode fibers, rotates a small portion of the MMF with respect to the rest, inducing controllable perturbations to linear mode couplings. The changes in output speckle shapes observed in Fig. 3a result from these differences in mode couplings.

Rotating the actuator by a set number of steps, each corresponding to 0.12°, at the beginning of every epoch allows quantitative analysis of its effect on the dataset's representation at the optical system's output, as shown in Fig. 3b. This analysis demonstrates that, without actuator movement, the system's characteristics remain stable in the experiment's timeframe. In contrast, induced perturbations increase the differences in speckle distributions linearly at smaller rates and before saturating at higher levels.

When these varying rates of externally induced perturbation are applied to the physical layer during the training, the online training method provides larger final accuracy advantages with faster drifts. Results in Fig. 3c indicate this advantage to reach 39% final accuracy improvement. Also, even without any active rotation, this experiment yields lower task accuracies compared to the experiment without the actuator in Fig. 2a. This reduction is due to high-order mode filtering caused by the tight winding of the MMF on the actuator, which has a bending diameter of 18 mm. This well-known effect induces higher losses for higher-order modes and couples them out of the fiber, observed with bending diameters starting from 40 mm [28]. The reduced task accuracy with fewer effective modes in the fiber underscores the importance of high dimensionality in physical layers.

# Effects of OLT Architecture on Modeling Fidelity and Computational Costs

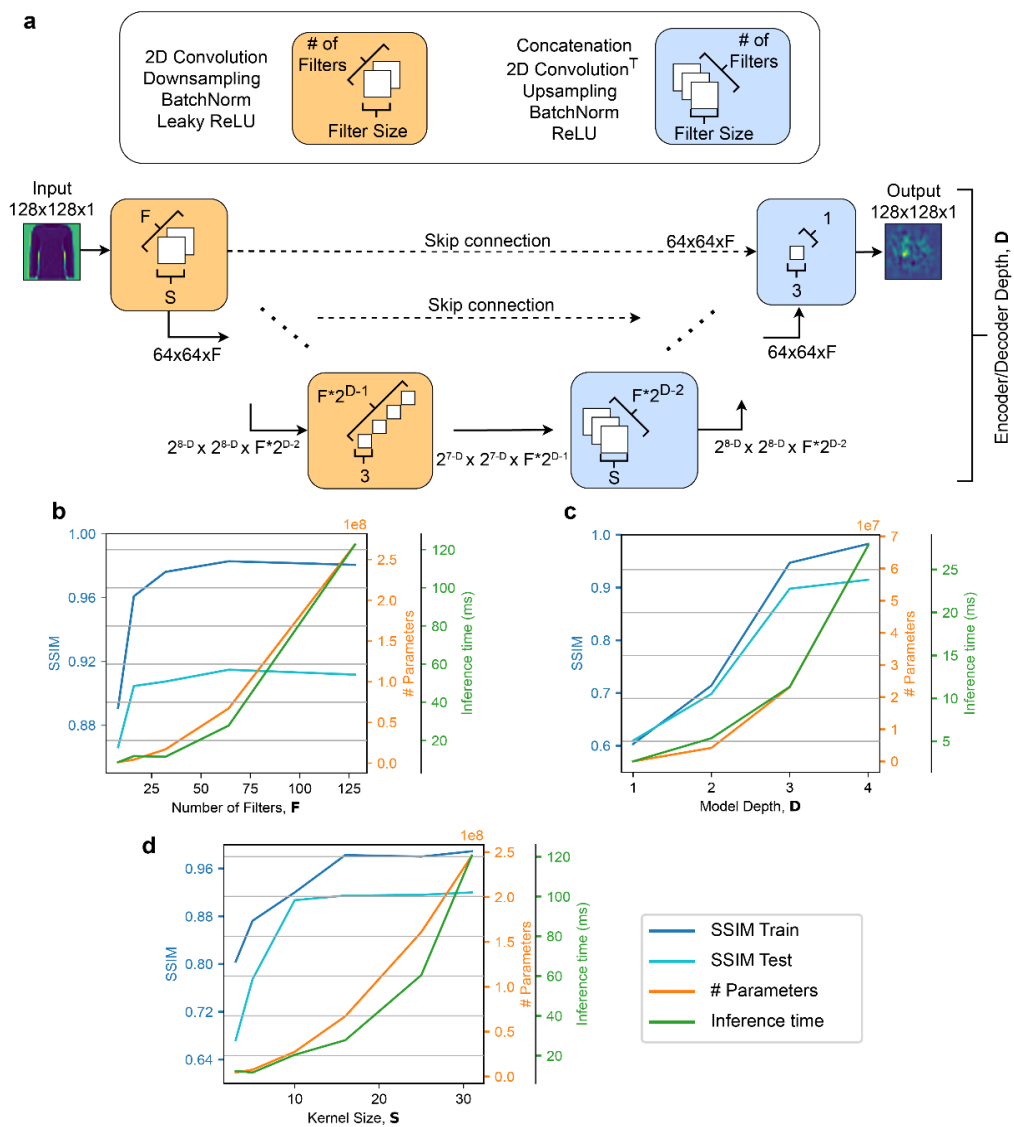

**Figure 4 Architecture of the OLT and its scaling behavior under various hyperparameter configurations. a** Network structure of the OLT, based on the U-Net architecture. The layers forming the downscaling (orange) and upscaling (blue) blocks, along with their dimensionalities, are shown. **b** Impact of the number of convolutional filters on structural similarity index (SSIM), total model parameters, and inference time per sample. **c** Effect of varying the number of downscaling and upscaling layers, which defines the model depth, on the same metrics. **d** Influence of convolutional kernel sizes on SSIM, model parameters, and inference time.

Capturing the complex dynamics of multimode nonlinear fiber optics with a neural network requires a large-scale, deep architecture. In this study, the mapping to be learned is from the two-dimensional input phase distribution on the SLM to the two-dimensional output intensity on the camera. A convolutional U-Net [29] architecture is particularly well-suited for this image-to-image translation task. It employs encoding blocks to downscale and extract features into a latent space, and decoding blocks to reconstruct the desired output from these latent features through upscaling, as visualized in Fig. 4a.

Optimizing the scale of this architecture for the task at hand requires careful analysis, as a smaller or disproportionate network might not be able to fully capture the underlying dynamics, while an overly complicated structure might tend to overfit or require excessive amount of compute resources. Using the dataset of inputs and outputs to the optical system, we trained different OLT architectures by varying the depth, defined by the number of encoding and decoding blocks. The improvement in the structural similarity index (SSIM) between OLT predictions and optical measurements, as shown in Fig. 4c, highlights the clear benefit of deeper architectures.

However, for architectures with 4 down- and up-sampling blocks, increasing the number of convolutional filters above 16 results in a significantly larger model and, more critically, longer inference times, without improving the accuracy, as demonstrated in Fig. 4b. Similarly, while increasing the size of convolutional kernels up to $16 \times 16$ improves accuracy, larger kernel sizes do not yield further gains, despite resulting in a much larger architecture as plotted in Fig. 4d.

## Conclusions

In this study, we studied an MMF-based optical system as a prototype for integrating complex, nonlinear physical phenomena as fixed layers into neural networks. To enable the training of such networks using the error backpropagation algorithm, which facilitated the artificial intelligence models to reach their current immense scales, a method for estimating the derivatives of the physical layer's outputs is required. This becomes crucial when the underlying physical relations are intractably complex for analytical computation.

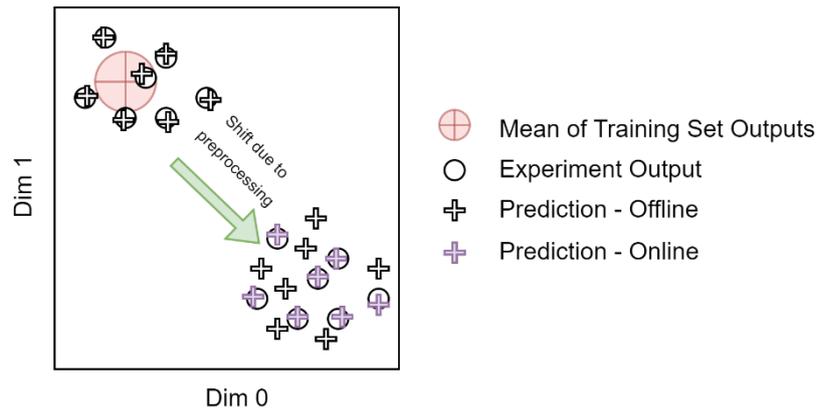

**Figure 5 The visualization of the changes in the optical system's and OLT outputs during the training process due to the updates to the preprocessing layers.**

Our approach employs an image-to-image translating neural network to model the dynamics of the optical system. This differentiable neural network approximates the system's outputs and also enables the calculation of gradients required for backpropagation. Through extensive exploration of hyperparameter configurations, we identify that a neural network with 70 million parameters, achieving an inference time of 30 ms on an NVIDIA RTX 3070 Ti GPU, provides the optimal trade-off between precision and computational cost, attaining an SNR of approximately 100. Furthermore, by continuously updating this proxy model, the method ensures accurate gradient predictions throughout the training procedure. Continuous updates to the model are particularly necessary when the distribution of inputs to the physical system shifts due to the weight updates in preceding layers. Fig. 5 illustrates the evolution of the optical system's and the OLT outputs over the training process. Notably, if the OLT is trained exclusively on raw sample inputs, changes in the input distribution can lead to reduced prediction accuracy. However, online learning effectively mitigates this issue, maintaining accuracy under evolving input distributions. While training with random inputs and their optical outputs can enhance the OLT's coverage, it reduces precision in specific subspaces due to the model's finite capacity.

Environmental and device-related factors can induce subtle changes in the transformation characteristics of analog systems. To investigate these effects, we artificially accelerated the system's drifts by incorporating a mechanical stage into the optical system, simulating systematic perturbations. Our findings reveal that while increased perturbations degrade neural network classification accuracy, updating the neural surrogate significantly mitigates this impact, improving task accuracy by nearly 40%.

This study demonstrates the symbiosis of a very large-scale optical and smaller scale digital layers in a neural network. Since the employed training algorithm, error backpropagation, has a strong track record in scaling up learning systems, the proposed approach paves the way for competitive, physics-assisted AI models with minimal digital computation needs in inference. For more complex tasks such as language or image generation, deeper architectures with multiple physical layers stacked in width and depth can be constructed using the principles outlined here.

We utilized optoelectronic devices to bridge physical layers with the digital domain and incorporate into neural networks, although physical interactions in optical data links, such as those in MMFs, are readily available and often treated as errors to be corrected. The proposed approach can reposition them as computational units by modeling these interactions with the OLT and incorporating them into neural networks. This enables existing optical data to be leveraged for computation, providing a pathway to fewer digital parameters and enhance AI efficiency, offering energy-efficient, low-latency workflows for environments like data centers.

# Appendix

## Appendix Note 1: Nonlinear Multimode Fiber Propagation

For an ideal fiber with a perfectly homogenous distribution of refractive index, light coupled to any mode propagates throughout on the same mode only with a phase change. However, perturbations to the ideal refractive index distribution and nonlinear propagation can cause the coupling of light from one mode to the other. This coupling can be formulized by using perturbation theory. The coupling coefficient is related to the overlap integral of the permittivity perturbation, $\Delta\epsilon(x,y,z)$, and the coupled mode shapes, $E_n(x,y)$, with a constant coefficient of $p$:

$$C_{p,n}(z) = p \int \Delta\epsilon(x,y,z) E_p^*(x,y) E_n(x,y) dx dy$$

In addition to the linear intermodal coupling due to waveguide perturbation, which simply can be induced by the bending of the fiber, nonlinear effects due to the light-matter interaction inside the waveguide can be represented as a mode-coupling mechanism. In general, these nonlinearities are represented as higher-order susceptibilities of the material and can be grouped as either parametric or non-parametric [30]. Parametric nonlinear processes do not change the quantum state of the material; hence they do not exchange energy with the material, and compared to non-parametric processes, they occur instantly. One of the most common optical parametric processes is the Kerr effect, manifesting itself in events such as self-phase modulation and cross-phase modulation, creating a phase shift and spectrum change [31]. On the other hand, non-parametric nonlinear optical processes create another set of effects on the propagation of intense light such as stimulated Raman and Brillouin scattering. In these events, the interaction of photons with matter creates and destroys phonons, changing photons' energy and momentum. Thus, through these processes, optical beams of different wavelengths are generated and amplified.

These linear and non-linear scattering events can be concisely formulated as coupling between modes of a multimode fiber. The following equation illustrates the evolution of each mode coefficient as it propagates through a multimode fiber by accounting for the effects of dispersion, linear coupling due to bending and other refractive index perturbations, and nonlinear coupling due to the Kerr effect and Raman scattering [32].

$$\frac{\partial A_p}{\partial z} = \underbrace{i\delta\beta_0^p A_p - \delta\beta_1^p \frac{\partial A_p}{\partial t} - i\frac{\beta_2}{2}\frac{\partial^2 A_p}{\partial t^2}}_{Dispersion} + \underbrace{i\sum_n C_{p,n} A_n}_{Linear\ Mode\ Coupling}$$

$$+ \underbrace{i\frac{\gamma}{3}\left(1 + \frac{i}{\omega_0}\frac{\partial}{\partial t}\right)\sum_{l,m,n} \eta_{plmn} \times \left[\underbrace{(1-f_R)A_l A_m A_n^*}_{Kerr\ contribution} + \underbrace{f_R A_l \int h_R(\tau) A_m(z,t-\tau) A_n^*(z,t-\tau) d\tau}_{Raman\ contribution}\right]}_{Nonlinear\ Mode\ Coupling}$$

Here, $\gamma$ is the nonlinear coefficient, $f_R$ is the Raman effect ratio in the nonlinear effect, $h_R(\tau)$ is the impulse response of the Raman scattering process and $\eta_{plmn}$ is the nonlinear mode coupling tensor, containing the amount of interaction for a set of 4 modes, and defined as:

$$\eta_{p,l,m,n} = \frac{\iint dxdy E_p E_l E_m E_n}{\left[\iint dxdy E_p \iint dxdy E_l \iint dxdy E_m \iint dxdy E_n\right]^2}.$$

In our experiments, we observed the ideal computing capabilities to occur at the level where maximum Kerr nonlinearities happen while Raman scattering is still negligible, which occurs around 10 kW of pulse peak power, as shown in Appendix Figure 1 [18]. For silica core, the first Raman peak is expected to be around 1080 nm for the pump around 1030 nm. The average fiber coupled power of 12.6 mW in this study is selected to satisfy this condition.

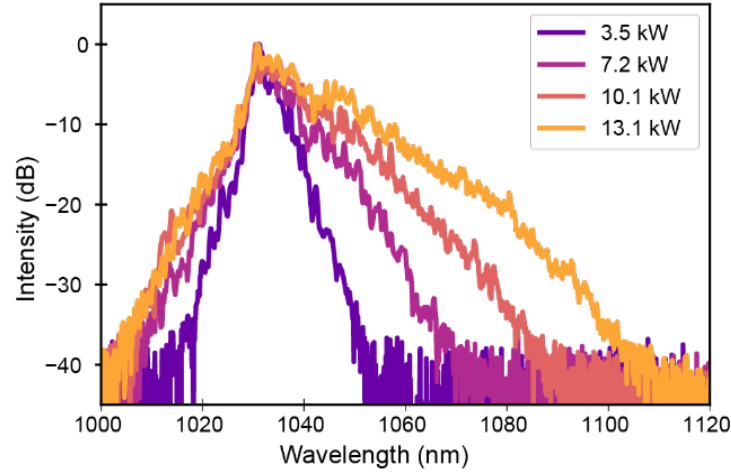

**Appendix Figure 1. Output spectra of the multimode fiber under different pulse peak power levels, reused from** [18].

## Appendix Note 2: Online Training Algorithm

In this methodology, the forward pass involves processing input $x$ through a series of layers and functions: the preprocessing layer $(f_p)$, the optical system $(f_{OS})$, the classifier $(f_c)$, and the loss function $(f_l)$. The overall operation can be expressed as: $(f_l \circ f_c \circ f_{OS} \circ f_p)(x)$. During the backward pass, the optical system $(f_{OS})$ is replaced by its differentiable proxy, the Optical Layer Twin $(f_{OLT})$, enabling gradients to flow back through the system.

$$\nabla(f_l \circ f_c \circ f_{OS} \circ f_p)(x) \approx J_{f_p}(x)^T J_{f_{OLT}}\left(f_p(x)\right)^T J_{f_c}\left(f_{OS}\left(f_p(x)\right)\right)^T \nabla f_l\left(f_c\left(f_{OS}\left(f_p(x)\right)\right)\right).$$

After gradient of the loss function is calculated, trainable layers are updated accordingly, for instance for the preprocessor, the update amount, $\Delta\theta_p$, is:

$$\Delta\theta_p \approx -\eta \nabla_{\theta_p}(f_l \circ f_c \circ f_{OS} \circ f_p)(x).$$

As it is also evident in this notation, the OLT's main functionality is to allow approximation of the Jacobian of the optical system such that $J_{f_{OLT}} \approx J_{f_{OS}}$. TensorFlow's *@custom_gradient* decorator allows the forward and backward computations to follow separate paths by using a single function, with the gradient of the experiment's output being explicitly defined. A key consideration is the use of vector-Jacobian products (VJP) instead of full Jacobian matrices, as the latter would be computationally prohibitive for high-dimensional data. For example, if the experiment maps a batch of 10 images of size $128 \times 128$ to equivalent outputs, the Jacobian will contain $(10 \cdot 128 \cdot 128)^2 \approx 1.7 \times 10^{10}$ elements. With 32-bit floating-point precision, this would require over 100 GB of memory, making it impractical for training. Instead, the VJP computes the gradient efficiently by combining the upstream gradient vector $u$ with the output of the OLT:

$$\text{custom\_gradient}(u) = \text{vjp}(u, f_{OLT}) = \nabla u \cdot f_{OLT} = u^T J_{f_{OLT}}.$$

At every step of training, the weights of the layers preceding and succeeding the optical layer (preprocessing and classifier layers) are updated following this formalism. In the "Online Learning" approach each step also refines the OLT by using the data obtained in the forward pass from the experiment. For the same inputs, let's call the output of the optical system as $y_{OS}$ and the output of the OLT as $y_{OLT}$, then the loss function for the refinement step can be defined as the squared error: $L_{\text{refine}} = |y_{OS} - y_{OLT}|^2$. The derivative of the loss function with respect to the parameters of the OLT provides the update direction and amount: $\Delta \theta_{OLT} = -\eta \frac{\partial L_{refine}}{\partial \theta_{OLT}}$, $\eta$ being the learning rate. This refinement step closely matches the behavior of the OLT to the experimental system, as the input distribution evolves due to updates in the preprocessing layer and possible physical drifts. This ensures high fidelity weight updates, reducing discrepancies and stabilizing the training process.

The presented experimental results are obtained with a preprocessing block of 6 linear convolutional layers of 1 kernel with $6 \times 6$ parameters, followed with a single sigmoid nonlinearity, and a classifying layer flattening $40 \times 40$ images and predicting the output with 10 softmax neurons. A stochastic gradient descent algorithm with $10^{-3}$ learning rate optimized the trainable parameters.